\documentclass[preprint,superscriptaddress,aps]{revtex4}

\usepackage{graphicx}
\usepackage{bm}

\newcommand{\beq}{\begin{equation}}
\newcommand{\eeq}{\end{equation}}
\newcommand{\beqa}{\begin{eqnarray}}
\newcommand{\eeqa}{\end{eqnarray}}

\newcommand{\w}{\omega}
\newcommand{\thetai}{\theta_i}

\begin{document}

\title{A new mechanism for electron spin echo envelope modulation}

\author{John~J.~L.~Morton}
\email{john.morton@materials.ox.ac.uk} \affiliation{Department of
Materials, Oxford University, Oxford OX1 3PH, United Kingdom}

\author{Alexei~M.~Tyryshkin}
\affiliation{Department of Electrical Engineering, Princeton
University, Princeton, NJ 08544, USA}

\author{Arzhang~Ardavan}
\affiliation{Clarendon Laboratory,
Department of Physics, Oxford University, Oxford OX1 3PU, United
Kingdom}

\author{Kyriakos~Porfyrakis}
\affiliation{Department of Materials, Oxford University, Oxford
OX1 3PH, United Kingdom}

\author{Stephen~A.~Lyon}
\affiliation{Department of Electrical Engineering, Princeton
University, Princeton, NJ 08544, USA}

\author{G.~Andrew~D.~Briggs}
\affiliation{Department of Materials, Oxford University, Oxford
OX1 3PH, United Kingdom}

\date{\today}

\begin{abstract}
Electron spin echo envelope modulation (ESEEM) has been observed
for the first time from a coupled \emph{hetero}-spin pair of
electron and nucleus in liquid solution. Previously, modulation
effects in spin echo experiments have only been described in
liquid solutions for a coupled pair of homonuclear spins in NMR or
a pair of resonant electron spins in EPR. We observe low-frequency
ESEEM (26 and 52~kHz) due to a new mechanism present for any
electron spin with $S>1/2$ that is hyperfine coupled to a nuclear
spin. In our case these are electron spin ($S=3/2$) and nuclear
spin ($I=1$) in the endohedral fullerene N@C$_{60}$. The
modulation is shown to arise from second order effects in the
isotropic hyperfine coupling of an electron and $^{14}$N nucleus.
\end{abstract}


\maketitle

\section{Introduction}
Measuring the modulation of a spin echo in pulsed magnetic
resonance experiments has become a popular technique for studying
weak spin-spin couplings. It is used extensively in the fields of
chemistry, biochemistry, and materials science, both in liquids
and solids, using nuclear magnetic resonance (NMR) and electron
paramagnetic resonance (EPR)
~\cite{ernst1987,dikanov1992,schweiger2001}. Two distinct
mechanisms for spin echo modulation have been identified in the
literature.

In the first mechanism, a pair of spins, $S$ and $I$, are coupled
through an exchange or dipole-dipole interaction,
$J\!\cdot\!\vec{S}\!\cdot\!\vec{I}$ or $J \!\cdot\! S_z I_z$, and the
echo modulation arises for non-selective refocusing pulses which flip
both coupled spins. The magnitude of the echo signal oscillates as
$\cos(Jt)$, where $t$ is the interpulse delay time
(\cite{abragam1961}, page~500). This is most commonly observed for
coupled pairs of homonuclear spins~\cite{hahn1952}, though it is also
known for pairs of coupled electron spins with identical or similar
Larmor frequencies~\cite{yudanov1969,milov1986}.

It should be emphasised that a similar coupling, $J \!\cdot\!  \vec{S}
\!\cdot\! \vec{I}$, between unlike spins (including coupling between
heteronuclear spins, hyperfine coupling between electron and nuclear
spins, and electron-electron coupling between two electron spins with
different Larmor frequencies) results in no modulation effects from
this mechanism: The hetero-spin coupling energy changes its sign upon
application of the refocusing pulse (because only one spin flips), and
the magnetization is thus fully refocused at the time of echo
formation, in the same way as in the presence of any other inhomogeneous
magnetic fields.

A second ESEEM mechanism, which does apply to coupled pairs of
hetero-spins, has also been
identified~\cite{dikanov1992,schweiger2001}. This mechanism
requires \emph{anisotropic} spin-spin interactions (e.g. $A_{zz}
S_z I_z + A_{zx} S_z I_x$) and is therefore restricted to solids
or high viscosity liquids. The modulation arises as a result of
"branching" of the spin transitions created by the refocusing
pulse. The resonant spin $S$ precesses with Larmor frequency that
is different before and after the refocusing pulse and therefore
accumulates an additional phase which causes oscillations in the
echo signal, as $\cos(\omega_{Ik} t)$, where $\omega_{Ik}$ is the
spin transition frequency of the non-resonant spin $I$. The
amplitude of the oscillations depends on magnitude of the
anisotropic hyperfine component.

In this Paper we demonstrate that, contrary to previous belief,
echo modulation effects can also be observed for a hetero-spin
pair coupled by a purely \emph{isotropic} spin interaction, and we
thus identify a new ESEEM mechanism. Our hetero-spin pair is the
endohedral fullerene N@C$_{60}$ in CS$_{2}$ solution, with
electron spin $S=3/2$ interacting through an isotropic hyperfine
coupling ($a\!\cdot\!\vec{S}\!\cdot\!\vec{I}$, $a=15.8$~MHz) to
the nuclear spin $I=1$ of $^{14}$N. The isotropic hyperfine
coupling lifts the degeneracy of the electron spin transitions,
leading to a profound modulation of the echo intensity at about
52~kHz. We shall show that this third modulation mechanism is only
effective in high-spin electron systems ($S>1/2$). The
N@C$_{60}$~molecule has an exceptionally long electron spin
dephasing time ($T_2=210$~$\mu$s),
 enabling the observation of this low frequency ESEEM for the first
time.

\section{Materials and Methods}
High-purity endohedral N@C$_{60}$ was prepared~\cite{mito},
dissolved in CS$_{2}$ to a final concentration of 1-2$\cdot
10^{15}$/cm$^3$, freeze-pumped in three cycles to remove oxygen,
and finally sealed in a quartz EPR tube. Samples were 0.7-1.4~cm
long, and contained approximately $5\cdot 10^{13}$ N@C$_{60}$
spins. Pulsed EPR measurements were done at 190~K using an X-band
Bruker Elexsys580e spectrometer, equipped with a nitrogen-flow
cryostat. In the 2-pulse (Hahn) electron spin echo (ESE)
experiments, $\pi/2-\tau-\pi-\tau-echo$, the $\pi/2$ and $\pi$
pulse durations were 56 and 112~ns respectively. Phase cycling was
used to eliminate the contribution of unwanted free induction
decay (FID) signals.

\section{Results and Discussion}

\begin{figure}[t]
\centerline {\includegraphics[width=3.2in]{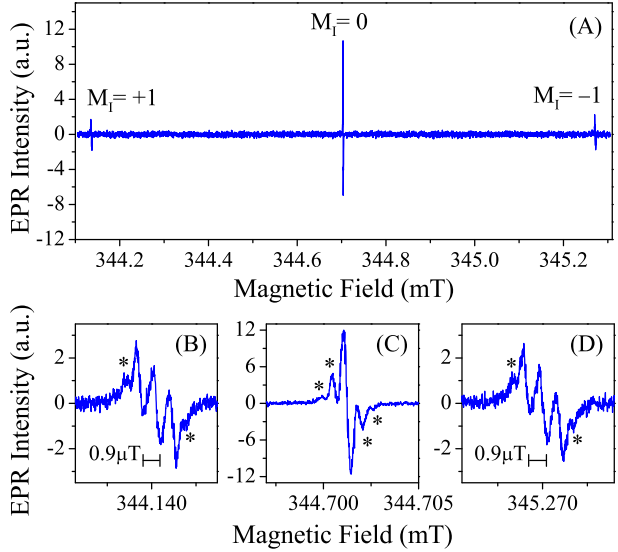}} \caption{(A)
EPR spectrum of N@C$_{60}$ in CS$_{2}$ at room temperature. Each
line in the triplet signal is labeled with the corresponding
projection $M_I$ of the $^{14}$N nuclear spin. (B-D) Zoom-in for
each line showing details of the lineshape structure. Small
satellite lines (marked with $\ast$) are due to a hyperfine
interaction with the natural abundance of $^{13}$C nuclei on the
C$_{60}$ cage. Measurement parameters: microwave frequency,
9.67~GHz; microwave power, 0.5~$\mu$W; modulation amplitude, 2~mG;
modulation frequency, 1.6~kHz.}\label{cwEPR}
\end{figure}

\begin{figure}[t]
\centerline {\includegraphics[width=3in]{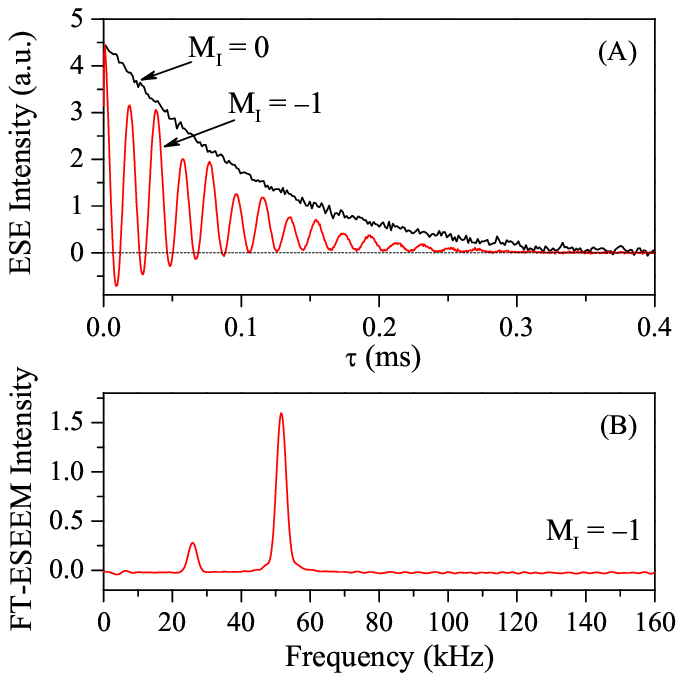}} \caption{(A)
Two-pulse ESE decays for N@C$_{60}$ in CS$_{2}$ at 190~K measured
at the central $M_I=0$ and the high-field $M_I=-1$ hyperfine
components of the EPR spectrum. (B) The Fourier Transform (FT) of
the oscillatory echo decay at $M_I=-1$.}\label{eseem}
\end{figure}

Fig.~\ref{cwEPR}(A) shows the continuous-wave EPR spectrum of
N@C$_{60}$ in CS$_{2}$ at room temperature. The spectrum is
centered on the electron g-factor $g=2.0036$ and comprises three
lines resulting from the hyperfine coupling to $^{14}$N
\cite{Murphy1996}. The relevant isotropic spin Hamiltonian (in
angular frequency units) is:
\begin{equation}\label{Hamiltonian}
\mathcal{H}_0=\w_e S_z - \w_I I_z + a \!\cdot\! \vec{S} \!\cdot\!
\vec{I},
\end{equation}
where $\w_e=g\beta B_0/\hbar$ and $\w_I=g_I\beta_n B_0/\hbar$ are
the electron and $^{14}$N nuclear Zeeman frequencies, $g$ and
$g_I$ are the electron and nuclear g-factors, $\beta$ and
$\beta_n$ are the Bohr and nuclear magnetons, $\hbar$ is Planck's
constant and $B_0$ is the magnetic field applied along $z$-axis in
the laboratory frame. Each hyperfine line (marked in
Fig.~\ref{cwEPR}(A) with $M_I=0$ and $\pm 1$) involves the three
allowed electron spin transitions $\Delta M_S=1$ within the
$S=3/2$ multiplet. These electron spin transitions remain
degenerate for $M_I=0$ as seen in Fig.~\ref{cwEPR}(C) but split
into three lines (with relative intensities 3:4:3) for $M_I=\pm
1$, as seen in Figs.~\ref{cwEPR}(B) and (D). This additional
splitting of 0.9~$\mu$T originates from the second order hyperfine
corrections $a^2/\w_e = 26$~kHz, and its observation is only
possible because of the extremely narrow EPR linewidth $<0.3~\mu$T
in N@C$_{60}$. Similar second-order splittings have been reported
for the related spin system of endohedral fullerene
$^{31}$P@C$_{60}$ which has $S=3/2$ coupled with $I=1/2$
\cite{Knapp1998}.

Fig.~\ref{eseem}(A) shows two-pulse echo decays measured at the
central $M_I=0$ and the high-field $M_I=-1$ hyperfine lines. The
decay is monotonic for $M_I=0$ and has an exponential dependence
$\exp(-2\tau/T_2)$ with $T_{2}=210~\mu$s. However, the decay is
oscillatory for $M_I=-1$ (and also for $M_I=+1$, not shown) ---
the Fourier transform of the decay reveals two peaks at
frequencies 26 and 52~kHz as seen in Fig.~\ref{eseem}(B). These
frequencies correlate closely to the splitting of 26~kHz found in
the EPR spectrum in Fig.~\ref{cwEPR}(B) and (D), indicating that
the two effects have the same origin.

We shall use the spin density operator formalism to derive the
modulation effects for the spin system $S=3/2$, $I=1$.  The spin
density matrix after our two-pulse echo experiment is given by

\beq \sigma(\tau)=(U_{\tau} R_2^x U_{\tau} R_1^x) \cdot \sigma_0
\cdot (U_{\tau} R_2^x U_{\tau} R_1^x)^\dag. \label{twopshort}\eeq

Here, $\sigma_0$ is the density matrix at thermal equilibrium; in
the high-temperature approximation valid in our experiments,
$\sigma_0$ can be substituted with a spin operator
$S_z$~\cite{schweiger2001}. The evolution operator
$U_{\tau}=\exp(-i\mathcal{H}_0 \tau)$ describes a free evolution
of the spin system between the applied microwave pulses, and the
spin rotation operators, $R_{i}^x$, describe spin rotation upon
application of the two microwave pulses, $i=1,2$. The measured
echo intensity is

\begin{equation}\label{twopexp}
V(\tau)=\mathrm{Tr}\left[ \sigma(\tau) \cdot D \right] .
\end{equation}

The detection operator $D=S_y \otimes \mathcal{P}_{M_I}$ involves
the $^{14}$N nuclear spin projection operator $\mathcal{P}_{M_I}$
to selectively detect only those spin transitions associated with
a specific nuclear spin projection $M_I$. In a pulsed EPR
experiment, this corresponds to performing measurements at the
resolved hyperfine line in the EPR spectrum of N@C$_{60}$ and
integrating over the echo signal shape to average out oscillating
signals from other off-resonance hyperfine lines. The echo derives
from a sum of the single quantum (SQ) coherences (represented by
the terms $\sigma_{n,n+1}$ and $\sigma_{n+1,n}$ in the density
matrix), weighted by factors from the detection operator $D$.

Our Hamiltonian, $\mathcal{H}_0$ (Eq.~\ref{Hamiltonian}), has
small off-diagonal elements provided by the
$a(I_xS_x+I_yS_y$)~terms. These terms are often omitted since they
are known to contribute only second order energy
corrections~\cite{abragam1961,slichter1996}, but are, in fact,
directly responsible for the observed ESEEM. Diagonalisation of
Eq.~\ref{Hamiltonian} yields the magnitude of these corrections to
be of order $\delta=a^2/\w_e$, consistent with the splitting
observed in Fig.~\ref{cwEPR}(B) and (D). In addition, the
isotropic hyperfine interaction introduces a small degree of
mixing between $I_z,S_z$ basis states, however we find that this
small mixing need not be considered to appreciate the origin of
the observed ESEEM. With this assumption, our Hamiltonian and all
other operators have block-diagonal structures with non-zero
elements only between states with the same $M_I$. Transitions with
simultaneous flip of both electron and nuclear spins are thus
forbidden~\cite{flipflop} and the evolution of electron spin can
be treated individually for each nuclear spin manifold. We can
therefore avoid the derivation in the full $12 \times 12$ Hilbert
space in a general form, and instead reduce the dimensionality to
$4 \times 4$. The validity of this approximation is confirmed
below by a rigorous derivation using Average Hamiltonian Theory.
The reduced $M_I=+1$ subspace of the diagonalised Hamiltonian,
correct to second order in $a$, becomes:

\begin{equation}\label{eq_h}
\mathcal{H}_0=S_z (\w_e+a)-I_z\w_I-
\left(%
\begin{array}{cccc}
  0  & 0 & 0 & 0 \\
  0 & \frac{3\delta}{2} & 0 & 0 \\
  0 & 0 & 2\delta & 0 \\
  0 & 0 & 0 & \frac{3\delta}{2} \\
\end{array}%
\right),
\end{equation}
which we can rearrange as:
\begin{equation}\label{eq_h2}
\mathcal{H}_0=S_z
(\w_e+a+\frac{\delta}{2})-I_z(\w_I+\frac{7\delta}{4})+
\left(%
\begin{array}{cccc}
  \delta  & 0 & 0 & 0 \\
  0 & 0 & 0 & 0 \\
  0 & 0 & 0 & 0 \\
  0 & 0 & 0 & \delta \\
\end{array}%
\right)I_z.
\end{equation}

The term $-I_z(\w_I+7/4\delta)$ represents a constant energy shift
and can be ignored. We move into a resonant rotating frame (the
coordinate system rotating with the microwave frequency
$\omega_{mw}$ around the laboratory $z$-axis). In this frame the
spin Hamiltonian~(\ref{eq_h}) transforms to $\mathcal{H} -
\omega_{mw} S_z$, such that:

\beq \label{h0}
\mathcal{H}_0=\left(%
\begin{array}{cccc}
  3/2\Delta+\delta & 0 & 0 & 0 \\
  0 & 1/2\Delta~~ & 0 & 0 \\
  0 & 0 & -1/2\Delta~~ & 0 \\
  0 & 0 & 0 & -3/2\Delta+\delta \\
\end{array}%
\right), \eeq

where $\Delta = \w_e + a + \delta/2 - \omega_{mw}$ is the
resonance offset frequency.

The rotation operator $R_{i}^x=\exp(-i(\mathcal{H}_0 +
\mathcal{H}_1) t_{p_i})$ can be simplified by taking into account
a finite excitation bandwidth of the microwave pulses, which are
selective upon one hyperfine line in the EPR spectrum.
$\mathcal{H}_1 = g\beta B_1 S_x/\hbar$ where $B_1$ is the
microwave magnetic field applied along $x$-axis in the rotating
frame and $t_{p_i}$ is the duration of the microwave pulses.
Furthermore, since the bandwidth of $B_1 \simeq 4.5$~MHz is large
compared to both the intrinsic EPR linewidth 9~kHz and the
second-order spitting 26~kHz of the outer lines in the EPR
spectrum, all three spin transitions within electron the $S=3/2$
multiplet are equally excited and the respective rotating operator
can be approximated as $R_{i}^x\simeq \exp(-i\mathcal{H}_1
t_{p_i})=\exp(-i \thetai S_x)$, where $\thetai=g\beta B_1
t_{p_i}/\hbar$ is the microwave pulse rotation angle. This results
in the following spin rotation operator for an on-resonance
hyperfine line:

\begin{widetext}
\begin{equation}\label{Rtheta}
R_{i}^x = \left(%
\begin{array}{cccc}
  \cos^3\frac{\thetai}{2} & \frac{i}{\sqrt{3}}(\sin^3\frac{\thetai}{2}+\sin\frac{3\thetai}{2}) & -\frac{1}{\sqrt{3}}(\cos^3\frac{\thetai}{2}-\cos\frac{3\thetai}{2}) & -i\sin^3\frac{\thetai}{2} \\
  \frac{i}{\sqrt{3}}(\sin^3\frac{\thetai}{2}+\sin\frac{3\thetai}{2}) & \frac{1}{3}(\cos^3\frac{\thetai}{2}+2\cos\frac{3\thetai}{2}) & -\frac{i}{3}(\sin^3\frac{\thetai}{2}-2\sin\frac{3\thetai}{2}) & -\frac{1}{\sqrt{3}}(\cos^3\frac{\thetai}{2}-\cos\frac{3\thetai}{2}) \\
  -\frac{1}{\sqrt{3}}(\cos^3\frac{\thetai}{2}-\cos\frac{3\thetai}{2}) & -\frac{i}{3}(\sin^3\frac{\thetai}{2}-2\sin\frac{3\thetai}{2}) & \frac{1}{3}(\cos^3\frac{\thetai}{2}+2\cos\frac{3\thetai}{2}) & \frac{i}{\sqrt{3}}(\sin^3\frac{\thetai}{2}+\sin\frac{3\thetai}{2}) \\
  -i\sin^3\frac{\thetai}{2} & -\frac{1}{\sqrt{3}}(\cos^3\frac{\thetai}{2}-\cos\frac{3\thetai}{2}) & \frac{i}{\sqrt{3}}(\sin^3\frac{\thetai}{2}+\sin\frac{3\thetai}{2}) & \cos^3\frac{\thetai}{2} \\
\end{array}%
\right).
\end{equation}
\end{widetext}

We need not consider the rotation operator for off-resonance
lines, as their excitation will lead to an oscillating echo signal
which is averaged out by selective detection (i.e. through the
detection operator $D$) and therefore does not contribute to the
overall echo signal.

We evaluate Eq.~\ref{twopshort} for the two-pulse echo experiment
$\pi/2-\tau-\pi-\tau-echo$, and rearrange terms for the purposes
of the discussion which follows:

\beq \sigma(\tau)=(U_{\tau} R_{\pi}^x U_{\tau})\cdot (R_{\pi/2}^x
\cdot \sigma_0 \cdot (R_{\pi/2}^x)^\dag)\cdot (U_{\tau} R_{\pi}^x
U_{\tau})^\dag. \eeq

Evaluating the heart of the echo sequence, $U_{\tau} R_{\pi}^x
U_{\tau}$, is instructive in understanding the source of the
observed modulation.  From Eq.~\ref{Rtheta}, a perfect $\pi$
rotation is:

 \beq \label{ppi} R_{\pi}^x=
\left(%
\begin{array}{cccc}
  0 & 0 & 0 & -i \\
  0 & 0 & -i & 0 \\
  0 & -i & 0 & 0 \\
  -i & 0 & 0 & 0 \\
\end{array}%
\right), \eeq

with the resulting echo sequence operator shown below.

\beq \label{echoexp} U_{\tau} R_{\pi}^x U_{\tau} =-i
\left(%
\begin{array}{cccc}
  0 & 0 & 0 & e^{2i\delta t} \\
  0 & 0 & 1 & 0 \\
  0 & 1 & 0 & 0 \\
  e^{2i\delta t} & 0 & 0 & 0 \\
\end{array}%
\right) \eeq

This indicates that over the course of the experiment, states
$S=\pm3/2$ pick up a phase of $2\delta t$ with respect to the
states $S=\pm1/2$, or in other words, the outer SQ coherences
($\sigma_{1,2},\sigma_{2,1},\sigma_{3,4}$ and $\sigma_{4,3}$)
oscillate with frequency 2$\delta$, while the phases of the inner
SQ coherences ($\sigma_{2,3}$ and $\sigma_{3,2}$) remain constant.

The initial SQ coherences are provided by the first rotation
($\pi/2$):

\beq R_{\pi/2}^x \cdot \sigma_0 \cdot (R_{\pi/2}^x)^\dag =\left(%
\begin{array}{cccc}
  0 & -i\sqrt{3}/2 & 0 & 0 \\
  i\sqrt{3}/2 & 0 & -i & 0 \\
  0 & i & 0 & -i\sqrt{3}/2 \\
  0 & 0 & i\sqrt{3}/2 & 0 \\
\end{array}%
\right), \label{initial}\eeq

and the measuring weighting factors for each coherence are:
$\sqrt{3}/2$, 1, and $\sqrt{3}/2$ (associated with the detection
operator $\mathcal{D}$). Together these imply that the three
coherences contribute to the measured echo intensity with relative
amplitudes 3:4:3. In other words, their sum will yield a constant
component, and one oscillating with frequency $2\delta$, with
respective amplitudes 4:6. This is confirmed upon evaluation of
Eq.~\ref{twopexp},
\begin{equation}\label{Vtau1_0}
V_{M_I=\pm1}(\tau)= 2 + 3\cos2\delta\tau,
\end{equation}
and is consistent with the observed echo in Fig.~\ref{eseem}B. The
modulation amplitude is deep and the echo signal can change its
sign at the minima.

\begin{figure}[t]
\centerline {\includegraphics[width=3.4in]{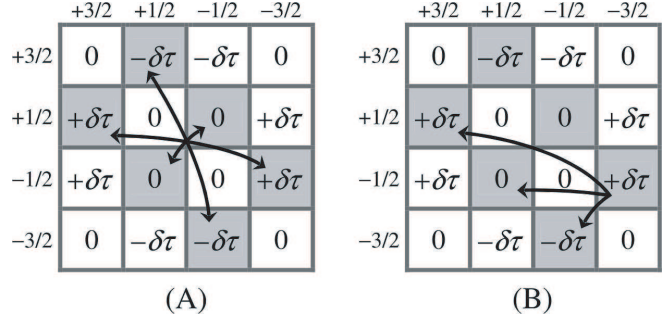}}
\caption{Phases gained by the density matrix elements during the
2-pulse ESE experiment at $M_I = +1$. Phases caused by
off-resonance refocus fully and are therefore omitted for clarity.
The shaded off-diagonal elements represent single-quantum (SQ)
coherences which generate the echo signal upon refocusing. In (A)
the arrows indicate the transition between the SQ coherences
caused by a perfect refocusing pulse with $\theta_2 = \pi$. (B)
shows all possible transitions (spin coherence branching), for one
SQ coherence element, caused by an imperfect refocusing pulse with
$\theta_2\neq \pi$.} \label{Dmatrix}
\end{figure}

The effect is
illustrated in Fig.~\ref{Dmatrix}(A), which shows the phases
gained during the ``defocusing'' period $\tau$, i.e. free
evolution after the initial pulse $\theta_1$ and before the
refocusing pulse $\theta_2$. These phases derive from the
differences between adjacent elements along the diagonal of the
Hamiltonian in Eq.~\ref{h0} (the off-resonance, $\Delta$, is
ignored as it is fully canceled upon echo formation). The six SQ
coherence elements, shaded in the Figure and responsible for echo
formation, gain the phases $0$ or $\pm \delta\tau$. Upon
application of the perfect refocusing pulse with $\theta_2 = \pi$
each SQ coherence element uniquely transforms, as shown with
arrows, and continues to evolve during the ``refocusing'' period
$\tau$ to gain an additional phase which does not compensate, but
instead doubles, the initial phase. Thus, at time of echo
formation the SQ coherences arrive with three different phases $0$
and $\pm 2\delta\tau$. Their vector sum interferes destructively
to produce an echo signal whose magnitude oscillates as
$2\delta\tau$ in accordance with Eq.~\ref{Vtau1_0}.

The case of an imperfect refocusing pulse with $\theta_2 \ne \pi$
is shown in Fig.~\ref{Dmatrix}(B). In contrast to $\theta_2 = \pi$
with a one-to-one transformation of each density matrix element,
the non-ideal pulse generates branching of the electron spin
transitions. Therefore, the SQ coherence element which initially
gained the phase $+\delta\tau$ during the ``defocusing'' period,
refocuses into three SQ coherences (shown with the arrows), each
accumulating different phases during the ``refocusing'' period.
Thus, at the time of echo formation the accumulated phases are
$0$, $+\delta\tau$, and $+2\delta\tau$. The vector sum of these
and other SQ coherences produces a complex interference with the
echo signal oscillating with two frequencies $\delta$ and
$2\delta$ as observed in Fig.~\ref{eseem}. Thus, the second
harmonic $\delta$ found in the echo modulation is the result of an
imperfect refocusing pulse.

The preceeding physical description provides an intuitive view of
the new ESEEM effect. For it to be fully rigorous, however, we
should consider the effect of the mixing of the $I_z,S_z$ basis
states. This necessarily involves the full 12-dimensional Hilbert
space and the argument rapidly becomes opaque. Fortunately, the
same results can be rigorously obtained from the first order
correction in Average Hamiltonian Theory (AHT)~\cite{waugh1968},
which is equivalent to a standard perturbation theory approach in
the rotating frame.

We begin by transforming the original Hamiltonian
(Eq.~(\ref{Hamiltonian})) into a rotating frame of angular
frequency $\omega_{mw}$, defining $\Omega_e$ as the deviation from
the electron Larmor frequency, $\Omega_e = \omega_e-\omega_{mw}$,
\begin{eqnarray}\label{Hrot}
\mathcal{H}_0 &{}={}& \Omega_e S_z - \w_I I_z + a [S_z I_z + S_x
I_x \cos(\w_{mw}t)\nonumber \\ &&\:{-} S_y I_y \sin(\w_{mw}t)].
\end{eqnarray}
The oscillatory terms in Eq.~\ref{Hrot} are averaged out in the
zeroth-order average Hamiltonian:
\begin{equation}\label{Haverage0}
\overline{\mathcal{H}}_0^{(0)} =\Omega_e S_z - \w_I I_z + a S_z
I_z.
\end{equation}
Here, the bar over $\mathcal{H}_0$ refers to an average over one
period of the oscillation $\w_{mw}$. As the Hamiltonian in
Eq.~(\ref{Haverage0}) results in no modulation of the echo signal
for a coupled hetero-spin pair, the higher order terms of the
average Hamiltonian must be included (see, for example,
Ref.~\cite{schweiger2001}, p.83). The first order correction is:
\begin{equation}\label{Haverage1}
\overline{\mathcal{H}}_0^{(1)} =\frac{\delta}{2}\left[ \left(
I(I+1)-I_z^2 \right)S_z - \left( S(S+1)-S_z^2 \right)I_z \right].
\end{equation}
We find that the average Hamiltonian $\overline{\mathcal{H}}_0 =
\overline{\mathcal{H}}_0^{(0)} + \overline{\mathcal{H}}_0^{(1)}$
is sufficient to describe the modulation effects. In this approach
the time-dependent mixing terms have been properly accounted for
to produce (after time averaging) the second order energy
corrections in $\overline{\mathcal{H}}_0^{(1)}$. However, these
mixing terms appear to average to zero, with the result that the
effective Hamiltonian, $\overline{\mathcal{H}}_0$, is a diagonal
matrix, thus validating our earlier qualitative approach.

It can also be verified that in the presence of the applied
microwave field, the average Hamiltonian is, to first order, the
simple sum $\overline{\mathcal{H}}_0 + g\beta B_1 S_x/\hbar$.
Therefore, the rotation operator, $R_{i}^x$, and detection
operator, $D$ have the same block diagonal structures described
above (non-zero elements only within the same $M_I$ subspace).
This allows us again to reduce the dimensionality of the Hilbert
space to $4 \times 4$.

Substituting Eqs.~(\ref{Rtheta}, \ref{Haverage0} and
\ref{Haverage1}) into Eq.~(\ref{twopexp}), and after some
manipulation, we find the following expressions for the echo
amplitude in a general two-pulse sequence
$\theta_1-\tau-\theta_2-\tau-echo$, for the $S=3/2, I=1$ spin
system. The echo modulation is identical for the two outer
$M_I=\pm 1$ hyperfine lines,
\begin{eqnarray}\label{Vtau1}
V_{M_I=\pm1}(\tau) &{}={}& 2\sin\theta_1\sin^2\frac{\theta_2}{2} [
A_0(\theta_2)+A_1(\theta_2)\cos\delta\tau \nonumber \\
&&\:{+}A_2(\theta_2)\cos2\delta\tau ],
\end{eqnarray}
 where
\begin{eqnarray}
A_0(\theta_2) &=&
1-6\cos^2\frac{\theta_2}{2}+\frac{27}{2}\cos^4\frac{\theta_2}{2}, \nonumber\\
A_1(\theta_2) &=& 6\cos^2\frac{\theta_2}{2}\left(%
2-3\cos^2\frac{\theta_2}{2} \right), \label{Acoeff}\\
A_2(\theta_2) &=& \frac{3}{2}\sin^2\frac{\theta_2}{2}\left(%
1-3\cos^2\frac{\theta_2}{2} \right), \nonumber
\end{eqnarray}
The signal is modulated with frequencies $\delta$ and $2\delta$,
consistent with the experimental observations in Fig.~\ref{eseem}.
The modulation amplitudes in Eq.~(\ref{Acoeff}) depend strongly on the
rotation angles of the microwave pulses. At optimal rotation angles,
$\theta_1 = \pi/2, \theta_2 = \pi$, Eq.~\ref{Vtau1_0} is recovered.

In contrast to the two outer lines, the echo signal at the central
$M_I=0$ hyperfine line shows no modulation effects,
\begin{equation}\label{Vtau0}
V_{M_I=0}(\tau)=2\sin\theta_1\sin^2\frac{\theta_2}{2}.
\end{equation}

It is instructive to consider which terms in the average
Hamiltonian $\overline{\mathcal{H}}_0^{(1)}$ give rise to the
modulation effects. The terms $I(I+1)S_z$ and $S(S+1)I_z$ are not
responsible as they produce only a constant shift to the electron
and nuclear Zeeman frequencies, respectively.  The term $I_z^2
S_z$ is also irrelevant because electron-nuclear flip-flop
transitions are forbidden, hence $M_I$ stays invariant during the
experiment. This term changes its sign, but not its magnitude,
during the refocusing pulse and thus fully refocuses. Therefore,
$S_z^2 I_z$ is solely responsible for the modulation effects. For
$M_I=0$, this last term becomes zero and consequently there are no
modulation effects produced at the central hyperfine line in the
EPR spectrum. Note that this term is responsible for the
\emph{spin-dependent} shifts to the energies of the spin states
used in the earlier derivation (see Eqs.~\ref{eq_h2}
and~\ref{h0}).

\begin{figure}[t]
\centerline {\includegraphics[width=3in]{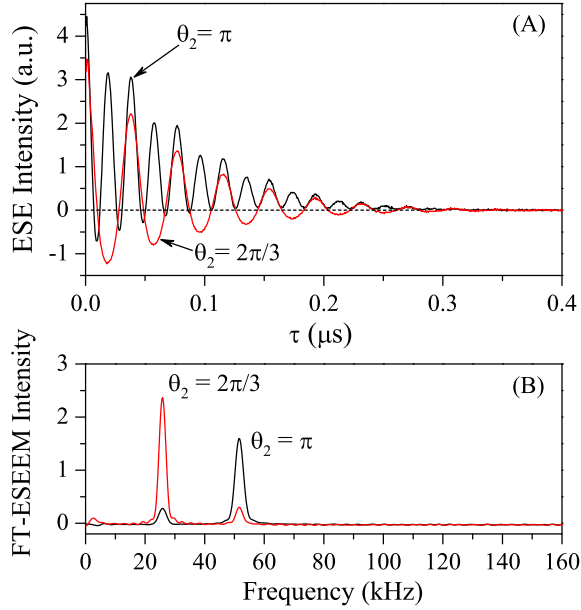}}
\caption{Two-pulse ESE decays (A) and their Fourier Transform
spectra (B) measured at hyperfine line $M_I=-1$ of the EPR
spectrum, using the refocusing pulse $\theta_2=\pi$ and
$\theta_2=2\pi/3$. Other experimental conditions are the same as
in Fig.~\ref{eseem}} \label{eseemNonIdeal}
\end{figure}

The effect of a non-$\pi$ refocusing pulse angle is shown in
Fig.~\ref{eseemNonIdeal}. As predicted by Eq.~(\ref{Vtau1}), we
observe that when $\theta_2 = 2\pi/3$, the modulation effects are
dominated by the low-frequency $\delta$, rather than the
high-frequency $2\delta$ found when $\theta_2 = \pi$.

Eq.~\ref{Vtau1} also confirms that a perfect $\pi$ refocusing
pulse yields only a $2\delta$ frequency component in the
modulation, however, a $\delta$ component is clearly observed in
Fig.~\ref{eseem}. The imperfection is explained by the
inhomogeneity of the microwave magnetic field $B_1$ in the
resonator cavity which results in a distribution of spin rotation
angles $\theta_2$ across the ensemble. If we assume a Gaussian
distribution of rotation angles, the relative intensities of the
low- and high-frequency components $I_1/I_2 = 0.17$ in the
experimental spectrum corresponds to a standard deviation of
$\sigma = 0.31$ radians. This corresponds to a 10\% error in a
$\pi$ rotation angle, consistent with previously reported values
for $B_1$-field inhomogeneity in this resonator cavity
\cite{Morton2004}. To verify that the $B_1$ field inhomogeneity is
the source of the low-frequency component, we applied an
error-correcting composite $\pi$-pulse as the refocusing pulse.
The resulting ESEEM contained only the single 52~kHz frequency
component.



The derivation of the modulation is easily generalised to the case
of an arbitrary electron spin $S > 1/2$ coupled through an
isotropic hyperfine interaction to a magnetic nucleus. Using an
approach similar to that described in
Eqs.~(\ref{ppi},~\ref{echoexp} and~\ref{initial}), the general
expression for 2-pulse ESEEM with a perfect refocusing pulse can
be shown to be
\begin{equation}\label{Vtau}
V(\tau)= \sum^{S}_{M_S=-S} (S-M_S)(S+M_S+1)\;
e^{i(1+2M_S)M_I\delta\tau}
\end{equation}
The summation is over electron spin projections $M_S$, whilst the
nuclear spin projection $M_I$ identifies the hyperfine line of the
EPR spectrum in which the modulation effects are observed.

\section{Conclusions}

Potential applications of this new mechanism include measuring the
hyperfine coupling constant and determining electron spin number;
it may also be relevant to certain quantum information processing
schemes~\cite{qipeseemnote}. The accurate measurement of the
hyperfine constant in a continuous-wave EPR measurement is subject
to $B_0$ field instability (typically $>$10~mG). However, the
ESEEM frequency can be measured accurately, given a sufficiently
long dephasing time, potentially providing a more precise
measurement. In this case accuracy may be improved by moving to
lower applied magnetic fields (lower EPR frequency). In contrast
with other types of ESEEM, this would also lead to a \emph{higher}
modulation frequency. Finally, we note that given appropriate
electron-nuclear spin coupling energies and decoherence times,
this effect will also lead to modulation in nuclear spin echo
experiments.

\section{Acknowledgements} We would like to thank Wolfgang
Harneit's group at the Hahn-Meitner Institute for providing
nitrogen-doped fullerenes, and John Dennis at Queen Mary's
College, London, Martin Austwick and Gavin Morley for the
purification of N@C$_{60}$. A Foresight LINK grant
\emph{Nanoelectronics at the quantum edge}, an EPSRC grant and the
Oxford-Princeton Link fund supported this project.  We thank
Brendon Lovett for valuable discussions. AA is supported by the
Royal Society. Work at Princeton was supported by the NSF
International Office through the Princeton MRSEC Grant No.
DMR-0213706 and by the ARO and ARDA under Contract No.
DAAD19-02-1-0040.


\begin{thebibliography}{15}
\bibitem[{fli()}]{flipflop}
\bibinfo{note}{These flip-flop transitions are allowed only to third order -
  the transition probability is proportional to $a^2/\w_e^3$ and thus is
  negligibly small at $a=15.8$~MHz for N@C$_{60}$.}
\bibitem[{qip()}]{qipeseemnote}
\eprint{The N@C$_{60}$ molecule has been proposed as an electron
spin-based
  qubit in several quantum information processing
  schemes~\cite{harneit,briggsRS}. At the very least, the slow evolution within
  the sublevels of the $S=3/2$ system which is responsible for the observed
  ESEEM must be taken into account when designing pulse sequences to perform a
  quantum algorithm. However, it could also be exploited to provide a separate
  family of gates for performing operations between sublevels, increasing the
  potential of N@C$_{60}$ as a single quantum bit.}
\end{thebibliography}
\end{document}